\documentclass[10pt,showpacs,english,twocolumn]{revtex4}
\usepackage{amsfonts,amsmath}
\usepackage{amssymb}
\usepackage{graphicx}
\usepackage{esint}

\newcommand{\be}{\begin{equation}}
\newcommand{\ee}{\end{equation}}
\newcommand{\ba}{\begin{eqnarray}}
\newcommand{\ea}{\end{eqnarray}}
\def\un{1\kern-3pt \rm I}

\def\I{\leavevmode\hbox{\small1\kern-3.8pt\norfeynmpmalsize1}}

\makeatletter
\makeatletter
\usepackage{babel}

\makeatother

\usepackage{babel}
\begin{document}

\title{Magnetic shielding of quantum entanglement states}

\author{O.M. Del Cima} \email{oswaldo.delcima@ufv.br} \author{D.H.T. Franco} \email{daniel.franco@ufv.br} \author{M.M. Silva} \email{marlon.silva@ufv.br} 
\affiliation{Universidade Federal de Vi\c{c}osa (UFV),\\ Departamento de F\'{\i}sica - Campus Universit\'ario,\\
Avenida Peter Henry Rolfs s/n - 36570-900 - Vi\c{c}osa - MG - Brazil.}

\begin{abstract}
The measure of quantum entanglement is determined for any dimer, either ferromagnetic or antiferromagnetic, spin-1/2 Heisenberg systems in the presence of external magnetic field. The physical quantity proposed as a measure of thermal quantum entanglement is the distance between states defined through the Hilbert-Schmidt norm. It has been shown that for ferromagnetic systems there is no entanglement at all. However, although under applied magnetic field, antiferromagnetic spin-1/2 dimers exhibit entanglement for temperatures below the decoherence temperature -- the one above which the entanglement vanishes. In addition to that, the decoherence temperature shows to be proportional to the exchange coupling constant and independent on the applied magnetic field, consequently, the entanglement may not be destroyed by external magnetic fields -- the phenomenon of {\it magnetic shielding effect of quantum entanglement states}. This effect is discussed for the  binuclear nitrosyl iron complex [Fe$_2$(SC$_3$H$_5$N$_2$)$_2$(NO)$_4$] and it is foreseen that the quantum entanglement survives even under high magnetic fields of Tesla orders of magnitude. 
\end{abstract}
\maketitle

\section{Introduction}

The quantum entanglement \cite{Amico,Guhne,Horodecki} is one of the most intriguing phenomenon in quantum physics. The first queries about its strangeness were contemporaneously raised by Einstein, Podolsky and Rosen \cite{Einstein35}, and by Schr\"odinger \cite{Schrodinger35}. 
Nowadays, the quantum entanglement is an important issue in quantum information theory \cite{Ekert91}, quantum teleportation \cite{Bennett00} and measurements techniques based on quantum information \cite{Raussendorf01}. Owing to its quantum nature, the entanglement could not be expected to emerge either at high temperatures or at larger scales than atomic. Nevertheless, experimental observations \cite{Ghosh03,Vertesi06,Rappoport07,Souza08} and theoretical proposals \cite{Gong2009,Mehran2014} of thermal entanglement in magnetic materials have been addressed so far in the literature. Furthermore, most recently experimental realizations in antiferromagnetic spin chains of long-distance entanglement and entanglement in surface-supported in presence of applied magnetic fields have been performed \cite{Sahling2015}. 

In this work, it has been investigated the dependence of quantum entanglement respecting to temperature and external (applied) magnetic fields for spin-1/2 Heisenberg dimer systems. As a physical quantity to measure quantum entanglement use has been made of the distance between states \cite{Vedral97,Vedral98} by adopting the Hilbert-Schmidt norm \cite{Dahl07}, together with the Peres-Horodecki criterion \cite{Peres96, Horodecki96}. The dependence of the quantum entanglement measure on temperature and applied magnetic field is analytically determined, add to that, it is verified that quantum entanglement becomes manifest just for antiferromagnetic spin-1/2 dimers -- no entanglement at all for ferromagnetic systems. Moreover, the decoherence (critical) temperature, above which the entanglement vanishes, is computed and, as a surprisingly by-product, it shows to be independent of the applied magnetic field, hence, the quantum entangled states are magnetically shielded -- the phenomenon of {\it magnetic shielding of quantum entanglement states} emerges. It should be stressed that the independence of the critical temperature with respect to the external magnetic field has also been tackled in earlier works \cite{Gong2009,Mehran2014}. Finally, analysing a  spin-1/2 antiferromagnetic spin-1/2 dimer [Fe$_2$(SC$_3$H$_5$N$_2$)$_2$(NO)$_4$], the binuclear nitrosyl iron complex \cite{nitrosyl}, it is foreseen that quantum entangled states resist applied magnetic fields larger than a Tesla order of magnitude.       

\section{The measure of quantum entanglement and the critical temperature}
\label{measure}
	
A dimer spin-1/2 Heisenberg system submitted to a constant external (applied) magnetic field can be described by the following Hamiltonian: 
\begin{equation}
\mathcal{H}=-J \vec S_1\cdot\vec S_2-g\mu_{\rm B}\sum_{i=1}^{2}\vec B\cdot\vec S_i~,
\label{HamHei1}
\end{equation}
where $J$ is the exchange coupling constant, $\vec S_i$ (i=1,2) are the spin operators, $\vec{B}$ is the external magnetic field applied to the system, $\mu_{\rm B}$ is the Bohr magneton and $g$ is the gyromagnetic factor. 

In order to express the Hamiltonian in matrix form, the spin operators are written in terms of Pauli matrices, $\vec S_1=\frac{1}{2}\vec\sigma_1=\frac{1}{2}\left(\sigma_{1}^{x},\sigma_{1}^{y},\sigma_{1}^{z}\right)$ and 
$\vec S_2=\frac{1}{2}\vec\sigma_2=\frac{1}{2}\left(\sigma_{2}^{x},\sigma_{2}^{y},\sigma_{2}^{z}\right)$, where the matrices $\vec \sigma$ are taken as 
\begin{subequations}
\label{Pau2x2}
\begin{align}
\vec\sigma_1&=\vec\sigma\otimes{\un}_2=\left(\sigma^{x}\otimes{\un}_2, \sigma^{y}\otimes{\un}_2,\sigma^{z}\otimes{\un}_2\right)~,\\
\vec\sigma_2&={\un}_2\otimes\vec\sigma=\left({\un}_2\otimes\sigma^{x}, {\un}_2\otimes\sigma^{y},{\un}_2\otimes\sigma^{z}\right)~,
\end{align}
\end{subequations}
with ${\un}_2\equiv{\un}_{2\times2}$ being the $2\times2$ identity matrix. 

Considering an applied magnetic field which is uniform and constant, pointing in $z$-direction, 
${\vec B}=B\hat{\rm\mathbf z}$, the Hamiltonian (\ref{HamHei1}) is rewritten as
\begin{equation}
\mathcal{H}=-\frac{J}{4} \vec\sigma_1\cdot\vec\sigma_2-\frac{1}{2}g\mu_{\rm B} B(\sigma_1^{z}+\sigma_2^{z})~,
\label{HamHei1sigma}
\end{equation}
therefore, 
\begin{eqnarray}
\label{MatHamDimHei}
\mathcal{H}=\left(
\begin{array}{cccc}
\displaystyle-\frac{J}{4}-g\mu_{\rm B} B & 0 & 0 & 0 \\
 0 & \displaystyle\frac{J}{4} & \displaystyle-\frac{J}{2} & 0 \\
 0 & \displaystyle-\frac{J}{2} & \displaystyle\frac{J}{4} & 0 \\
 0 & 0 & 0 & \displaystyle-\frac{J}{4}+g\mu_{\rm B} B \\
\end{array}
\right)~.
\end{eqnarray}

A finite temperature thermal equilibrium state is represented by the density operator $\rho={\mathcal Z}^{-1}{e^{-\beta {\mathcal H}}}$, where ${\mathcal Z}={\rm Tr}(e^{-\beta {\mathcal H}})$ is the partition function and $\beta=(k_{\rm B}T)^{-1}$. Taking into account the Hamiltonian $\mathcal{H}$ (\ref{MatHamDimHei}), which represents a dimer spin-1/2 Heisenberg system subjected to an uniform and constant external magnetic field along the $z$-direction, the density matrix $\rho$ is now written as 
\begin{eqnarray}
\label{MatDenDim}
\rho=\left(
\begin{array}{cccc}
 \rho _{11} & 0 & 0 & 0 \\
 0 & \rho _{22} & \rho _{23} & 0 \\
 0 & \rho _{32} & \rho _{33} & 0 \\
 0 & 0 & 0 & \rho _{44} \\
\end{array}
\right)~,
\end{eqnarray}
where
\begin{subequations}
\label{EleMatDenDimComCam}
\begin{align}
\rho _{11}&=\frac{1}{\mathcal Z}e^{\frac{\mathcal J}{4 T}+\frac{\mathcal B}{T}}~,\\ 
\rho _{22}=\rho _{33}&=\frac{1}{\mathcal Z}e^{-\frac{\mathcal J}{4 T}} \cosh \left(\frac{\mathcal J}{2 T}\right)~,\\
\rho _{23}=\rho _{32}&=\frac{1}{\mathcal Z}e^{-\frac{\mathcal J}{4 T}} \sinh \left(\frac{\mathcal J}{2 T}\right)~,\\
\rho _{44}&=\frac{1}{\mathcal Z}e^{\frac{\mathcal J}{4 T}-\frac{\mathcal B}{T}}~,
\end{align}
\end{subequations}
with the partition function ${\mathcal Z}\equiv\mathcal{Z}({\mathcal J},{\mathcal B},T)$ given by
\begin{eqnarray}
\label{FunParDimComCam}
\mathcal{Z}({\mathcal J},{\mathcal B},T)=e^{-\frac{4 \mathcal B+3 \mathcal J}{4 T}} \left(e^{\frac{\mathcal B+\mathcal J}{T}}+e^{\frac{2 \mathcal B+\mathcal J}{T}}+e^{\frac{\mathcal B}{T}}+e^{\frac{\mathcal J}{T}}\right)~,
\end{eqnarray}
where $\mathcal J=J/k_{\rm B}$ e $\mathcal B=g\mu_{\rm B} B/k_{\rm B}$.

\subsection{The Peres-Horodecki criterion} 

In order to call forth the Peres-Horodecki criterion \cite{Peres96,Horodecki96}, the partial transpose (T$_{\rm P}$) of the density matrix (\ref{MatDenDim}) has to be computed, which consists just by taking the transpose (T) in respect only to one of the subspaces -- the Hilbert subspaces $\mathbb H_1$ and $\mathbb H_2$ -- of the system and thus realizing the tensor product $\mathbb H_1 \otimes \mathbb H_2^{\rm T}$. The partial transpose relative to the subspace $\mathbb H_2$ reads
\begin{eqnarray}
\label{MatDenDimTraPar}
\rho^{{\rm T}_{\rm P}}=\left(
\begin{array}{cccc}
 \rho _{11} & 0 & 0 & \rho _{23} \\
 0 & \rho _{22} & 0 & 0 \\
 0 & 0 & \rho _{33} & 0 \\
 \rho _{32} & 0 & 0 & \rho _{44} \\
\end{array}
\right)~.
\end{eqnarray}
The Peres-Horodecki criterion \cite{Peres96,Horodecki96} ensures that, for $2\otimes2$ and $2\otimes3$ dimensional systems, the positivity of the partial transpose eigenvalues is a necessary and sufficient  condition for separability of quantum mixed states. 

The eigenvalues of the partial transpose, $\rho^{{\rm T}_{\rm P}}$, follow
\begin{subequations}
\label{AutValTraParDim}
\begin{align}
\lambda_1&=\rho _{22}~,\\
\lambda_2&=\rho _{33}~,\\
\lambda_3&=\frac{1}{2} \left(\rho _{11}+\rho _{44}+\sqrt{(\rho _{11}-\rho _{44})^2+4 \rho _{23} \rho _{32}}\right)~,\\
\lambda_4&=\frac{1}{2} \left(\rho _{11}+\rho _{44}-\sqrt{(\rho _{11}-\rho _{44})^2+4 \rho _{23} \rho _{32}}\right)~.\label{lambda4}
\end{align}
\end{subequations}
Bearing in mind the density matrix elements (\ref{EleMatDenDimComCam}) and the eingenvalues (\ref{AutValTraParDim}) all together, it is straightforwardly verified that the first three eigenvalues, $\lambda_1$, $\lambda_2$ and $\lambda_3$, are positive definite for any value, positive or negative, of the exchange coupling constant ($\mathcal J=J/k_{\rm B}$). However, the last eigenvalue, $\lambda_4$, is positive ($\lambda_4>0$) at any temperature provided that $\mathcal J>0$ (ferromagnetic), on the other hand, it is negative ($\lambda_4<0$)  below a critical temperature whenever $\mathcal J<0$ (antiferromagnetic). Consequently, there is no quantum entangled states, thus only separable, for dimer spin-1/2 ferromagnetic ($\mathcal J>0$) systems at any finite temperature, since $\lambda_4>0$ anyhow. Nevertheless, for dimer spin-1/2 antiferromagnetic ($\mathcal J<0$) systems quantum entanglement could emerge provided there is a critical (decoherence) temperature ($T_{\rm c}$) below which ($T<T_{\rm c}$) the eingenvalue $\lambda_4<0$, showing up in this case quantum entangled states, otherwise whenever at, and above, the decoherence temperature ($T\geq T_{\rm c}$) the entanglement should vanish, namely the eigenvalue $\lambda_4\geq 0$.        

The decoherence temperature is defined as the critical temperature below which the system exhibits quantum entanglement, the other way about, whenever the system reaches the critical temperature, or stand above it,  the quantum entanglement does not take place. Adding to that, in the case of dimer spin-1/2 antiferromagnetic ($\mathcal J<0$) systems, from the Peres-Horodecki criterion \cite{Peres96,Horodecki96}, the density matrix elements (\ref{EleMatDenDimComCam}) and the eingenvalues (\ref{AutValTraParDim}), the critical temperature ($T_{\rm c}$) can be determined by assuming that the forth eigenvalue, $\lambda_4\equiv\lambda_4(\mathcal J,\mathcal B,T$), vanishes at $T=T_{\rm c}$, thus  $\lambda_4(\mathcal J,\mathcal B,T_{\rm c})=0$, which results:
\begin{eqnarray}
\label{TemCriDim2}
T_{\rm c}=-\frac{\mathcal J}{\ln(3)}~,
\end{eqnarray}
for any component value of the external ($z$-direction) magnetic field. It shall be drawn the attention to the fact that in despite of the dimer spin-1/2 antiferromagnetic system be subjected to an external magnetic field, the critical temperature (\ref{TemCriDim2}) remains unaffected by the presence of the magnetic field, hence it follows that the quantum entanglement is magnetically shielded, it cannot be destroyed by the exposal to magnetic fields -- conjectured as the phenomenon of magnetic shielding of quantum entanglement states at finite temperatures.

\subsection{The measure of quantum entanglement: distance between states}

\begin{figure}[t]
\includegraphics[scale=0.35]{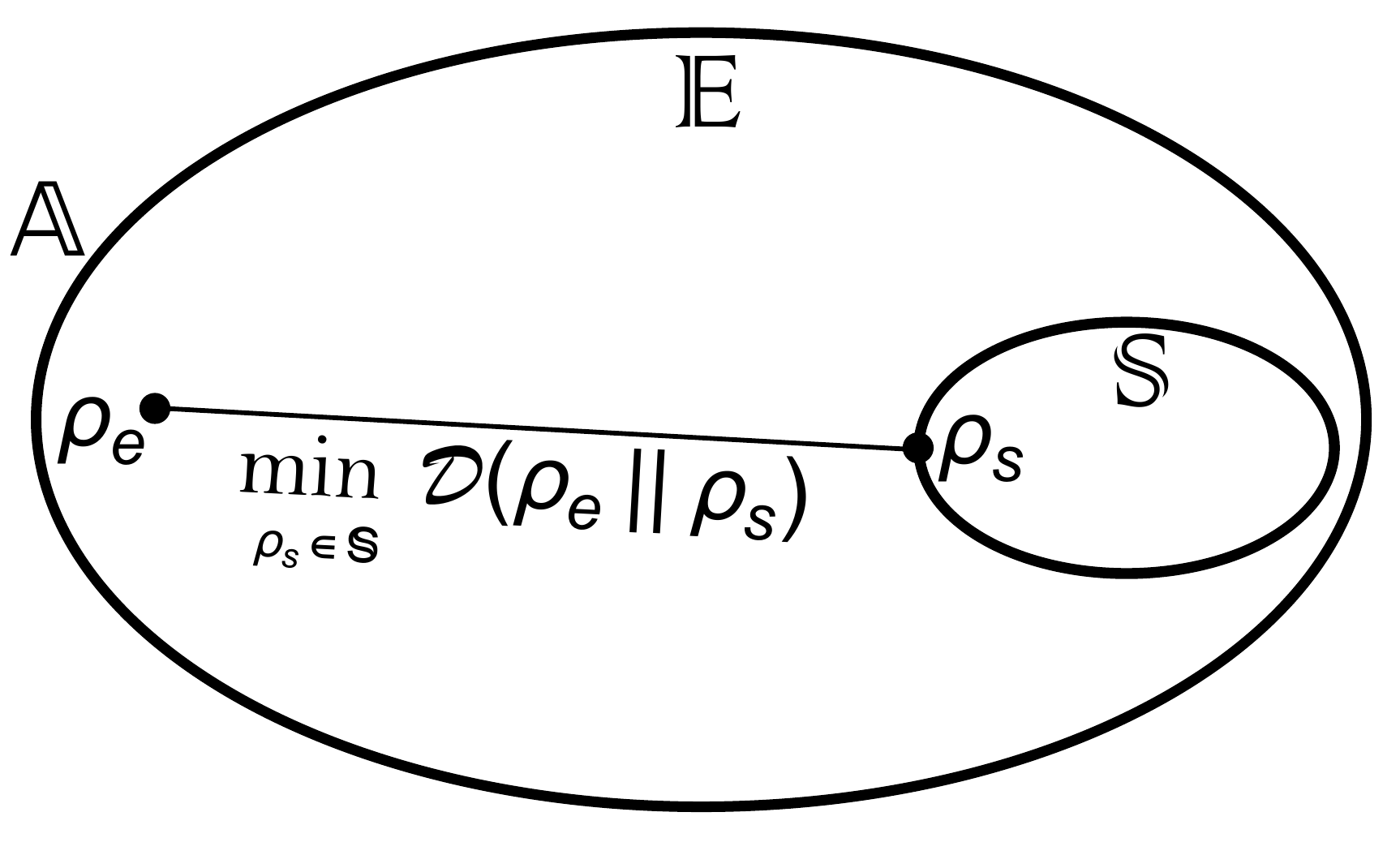}
 \caption{
The minimum distance among $\rho_e$ and $\rho_s$ states.}
\label{DesenhoDosEstados} 
\end{figure}

The physical quantity proposed here to quantify the thermal quantum entanglement is the distance between states, as introduced in \cite{Vedral97,Vedral98}, nevertheless the distance is defined through the Hilbert-Schmidt norm \cite{Dahl07}. Let $\mathbb A$ the set of all matrices densities based upon the tensor product of two Hilbert spaces $\mathbb H_1 \otimes\mathbb H_2$, which consists of two disjoints subsets: the subset of separable states, $\mathbb S$, and the subset of entangled states, $\mathbb E=\mathbb A-\mathbb S$. By definition, the proposed measure of entanglement ($\mathcal{E}$) is a physical quantity which lies in the internal $0\leq\mathcal E\leq1$ and is defined by:
\begin{equation}
\label{MedEmaDim}
\mathcal{E}(\rho_e)=\mathcal{E}_0 \min_{\rho_s\in\mathbb S} \mathcal D(\rho_e||\rho_s)~,
\end{equation}
where $\mathcal D(\rho_e||\rho_s)$ is the distance between the entangled state, $\rho_e \in \mathbb E=\mathbb A-\mathbb S$, and the separable state, $\rho_s \in \mathbb S$, with $\mathbb A$ and $\mathbb S$ being strictly convex sets (Fig.\ref{DesenhoDosEstados}), implying that every entangled state $\rho_e \in \mathbb E$ contains a unique best approximation to $\rho_e$ in $\mathbb S$, since all non-empty closed convex set is a Chebyshev set \cite{Bauschke10}. Also, the positive parameter $\mathcal{E}_0$ in (\ref{MedEmaDim}) is a normalization constant so as to ensure that the condition $0\leq\mathcal E(\rho_e)\leq1$ be satisfied.

Bearing in mind the Peres-Horodecki criterion \cite{Peres96,Horodecki96}, a candidate for the separable state density matrix, $\rho_s$, is allowed if its all partial transpose eigenvalues be positive definite, on the other hand, to make a choice of a candidate to the entangled state density matrix, $\rho_e$, at least one of its partial transpose eigenvalues must be negative. As previously mentioned, among the four eigenvalues (\ref{AutValTraParDim}), the only one which can be negative is $\lambda_4$ (\ref{lambda4}), consequently, the following inequalities stem from (\ref{EleMatDenDimComCam}) and (\ref{lambda4}):
\begin{subequations}
\label{ConEmaSepDim}
\begin{align}
\label{ConEmaDim}
(\rho_{23}^e)^2& > \rho_{11} \rho_{44}~,\\
\label{ConSepDim}
(\rho_{23}^s)^2& \leq \rho_{11} \rho_{44}~,
\end{align}
\end{subequations}
with the subscript-superscript indices $e$ and $s$ referring to entangled and separable state conditions, respectively. Therefore, the density matrix $\rho$ (\ref{MatDenDim}) can be rewritten as 
\begin{eqnarray}
\rho_{I}=\left(
\begin{array}{cccc}
 \rho _{11} & 0 & 0 & 0 \\
 0 & \rho _{22} & \rho _{23}^{I} & 0 \\
 0 & \rho _{23}^{I} & \rho _{33} & 0 \\
 0 & 0 & 0 & \rho _{44} \\
\end{array}
\right),
\end{eqnarray}
where $I=\{e,s\}$, such that $\rho_{23}^e$ is the element of the entangled state density matrix satisfying the condition (\ref{ConEmaDim}), whereas $\rho_{23}^s$ is the element of the separable state density matrix fulfilling the condition (\ref{ConSepDim}). 

As already mentioned, it has been proposed the Hilbert-Schmidt norm \cite{Dahl07} to compute the distance among an entangled and separable states, $\rho_e$ and $\rho_s$, respectively, since it satisfies necessary requirements for an entanglement measure \cite{Vedral98,Witte99}. 
The distance $\mathcal D(\rho_e||\rho_s)$ among entangled and separable states, defined through the Hilbert-Schmidt norm, follows
\begin{eqnarray}
\label{DRhosRhoe}
\mathcal D(\rho_s||\rho_e)&=&\sqrt{{\rm Tr}[(\rho_s-\rho_e)^2]} \nonumber\\
&=&\sqrt{2}|\rho_{23}^e-\rho_{23}^s|~.
\end{eqnarray}
Nevertheless, the proposed entanglement measure $\mathcal{E}(\rho_e)$ (\ref{MedEmaDim}) written now as 
\begin{equation}
\label{MedEmaDim1}
\mathcal{E}(\rho_e)=\mathcal{E}_0 \sqrt{2} \min_{\rho_s\in\mathbb S}|\rho_{23}^e-\rho_{23}^s|~,
\end{equation}
shall be a minimum as well as the conditions (\ref{ConEmaSepDim}) have to be adopted, reminding that for dimer spin-1/2 antiferromagnetic systems, thus $\mathcal J<0$, the density matrix elements $\rho_{11}$ and $\rho_{44}$ are positive, whereas $\rho_{23}$ is negative. Consequently, the condition (\ref{ConEmaSepDim}) can be rewritten as  
\begin{subequations}
\label{ConEmaSepDim2}
\begin{align}
\label{ConEmaDim2}
\rho _{23}^e<-\sqrt{\rho _{11} \rho _{44}}&\\
\label{ConSepDim2}
-\sqrt{\rho _{11} \rho _{44}}&\leq\rho _{23}^s<0~.
\end{align}
\end{subequations}

The distance defined by means of the Hilbert-Schmidt norm, $\mathcal D(\rho_s||\rho_e)$ (\ref{DRhosRhoe}), reaches a minimum when $\rho_{23}^{s}$ assumes its minimum value,  $\rho_{23}^{s}=-\sqrt{\rho_{11}\rho_{44}}$, and owing to $\rho_{23}<0$ ($\mathcal J<0$), the measure of entanglement ($\mathcal{E}(\rho_e)$) (\ref{MedEmaDim}) is redefined by  
\begin{equation}
\label{MedEmaDimComCam1_e}
\mathcal{E}(\rho_e)=-\mathcal{E}_0\sqrt2\left(\rho_{23}^e+\sqrt{\rho_{11}\rho_{44}}\right)~,
\end{equation}
where it shall be assumed $\mathcal{E}_0=\sqrt2$ in (\ref{MedEmaDimComCam1_e}) so as to guarantee that $0\leq\mathcal{E}(\rho_e)\leq1$, in which the extremes of that interval, namely for $\mathcal{E}(\rho_e)=0$ the system is disentangled while for $\mathcal{E}(\rho_e)=1$ it is maximally entangled. Nevertheless, for any state $\rho$, be it entangled ($\rho_e$) or separable ($\rho_s$), recalling the conditions (\ref{ConEmaSepDim2}), a general expression yields up:
\begin{equation}
\label{MedEmaDimComCam1}
\mathcal{E}(\rho)=\max\left[0,-2\left(\rho _{23}+\sqrt{\rho_{11}\rho_{44}}\right)\right]~.
\end{equation}

At this moment, by taking into consideration the density matrix elements (\ref{EleMatDenDimComCam}), the measure of entanglement  (\ref{MedEmaDimComCam1}) -- $\mathcal{E}(\rho)\equiv\mathcal{E}({\mathcal J},{\mathcal B},T)$ -- reads 
\begin{eqnarray}
\label{MedEmaDimComCam2}
&&\mathcal{E}({\mathcal J},{\mathcal B},T)= \nonumber\\
&&=\max\left[0,\frac{e^{\frac{\mathcal B}{T}} \left(1-3 e^{\frac{\mathcal J}{T}}\right)}{e^{\frac{\mathcal B+\mathcal J}{T}}+e^{\frac{2 \mathcal B+\mathcal J}{T}}+e^{\frac{\mathcal B}{T}}+e^{\frac{\mathcal J}{T}}}\right]~,
\end{eqnarray}
whence, a similar result has been obtained via concurrence \cite{Vedral01}. In the absence of external magnetic field ($\mathcal B=0$), the measure of entanglement (\ref{MedEmaDimComCam2}) recovers the result obtained by concurrence \cite{Aldoshin14}. However, it should be stressed that the entanglement measure (\ref{MedEmaDim}) proposed here, wherein the distance among states is defined through the Hilbert-Schmidt norm (\ref{DRhosRhoe}), may be used to measure entanglement in $2\otimes3$ dimensional systems -- such as mixed spin-(1/2,1) dimers \cite{spin-1/2-1} -- where the concurrence cannot be analytically computed. 

Whenever the system be in absence (${\mathcal B=0}$) or presence (${\mathcal B\neq0}$) of external magnetic field, the measure of entanglement (\ref{MedEmaDimComCam2}) can be rewritten by means of an unique expression which takes care either of the both situations: 
\begin{eqnarray}
\label{RelEntDis}
\mathcal{E}({\mathcal J},{\mathcal B},T)=\max\left[0,\frac{e^{-\frac{3 \mathcal J}{4 T}} \left(1-3 e^{\frac{\mathcal J}{T}}\right)}{\mathcal{Z}({\mathcal J},{\mathcal B},T)}\right],
\end{eqnarray}
with $\mathcal{Z}({\mathcal J},{\mathcal B},T)$ being the partition function given by (\ref{FunParDimComCam}) -- valid for vanishing and non vanishing applied magnetic field, whence for the former case it suffices to set ${\mathcal B=0}$ within (\ref{FunParDimComCam}). At this moment, it is propitious to notice that the decoherence temperature ($T_{\rm c}$) is defined as the temperature above which the quantum entanglement vanishes, therefore $T=T_{\rm c}$ implies that (\ref{RelEntDis}) becomes null, $\mathcal{E}({\mathcal J},{\mathcal B},T_{\rm c})=0$, being its solution precisely that one (\ref{TemCriDim2}) deduced from the Peres-Horodecki criterion. Such a congruence among the Peres-Horodecki criterion and the entanglement measure proposed here (\ref{RelEntDis}) -- expressed through the distance among states defined by the Hilbert-Schmidt norm -- makes manifest that it shall be a good quantum entanglement measure.

\section{The quantum entanglement magnetic shielding effect} 
\label{shielding}
 
The quantum entanglement measure (\ref{RelEntDis}) depends on three parameters: one intrinsic to the system, the exchange coupling constant (${\mathcal J}$); and two extrinsic, the temperature ($T$) and the external magnetic field (${\mathcal B}$). A detailed analysis concerning the proposed quantum entanglement measure (\ref{RelEntDis}) and its dependence with respect to the extrinsic physical quantities, temperature and  applied magnetic field, will be presented following.

\subsection{Entanglement dependence on temperature}

The behaviour of the entanglement measure (\ref{RelEntDis}) with respect to the temperature can be split into  three external magnetic field regimes in comparison by  with the exchange coupling constant: weak field ($|\mathcal B|<|\mathcal J|$), medium field ($|\mathcal B|=|\mathcal J|$) and strong field ($|\mathcal B|>|\mathcal J|$). Whatsoever the regime may be, the critical (decoherence) temperature is independent of the applied magnetic field and, in addition to that, though the system be subjected or not to an external magnetic field, the greater the absolute value of the exchange coupling constant, the greater is the decoherence temperature, thus it could be conjectured that the exchange coupling protects the system from decoherence as temperature increases up to it reaches the critical one and the quantum entanglement vanishes, consequently. Moreover, the stronger the applied magnetic field, the weaker the quantum entanglement -- the smaller the value of entanglement measure (\ref{RelEntDis}) -- for a given temperature.

\begin{figure}[t!]
\includegraphics[scale=0.5]{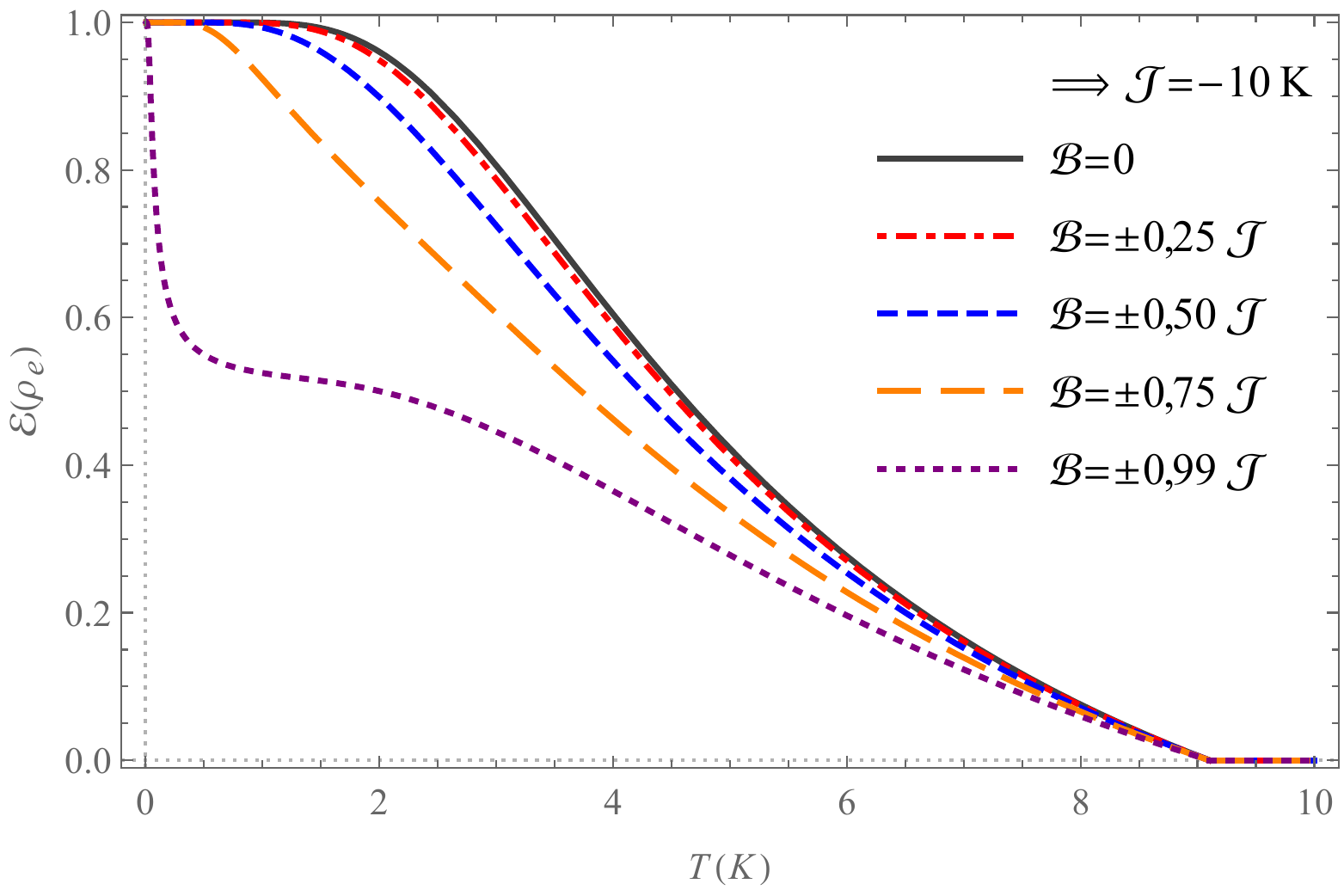}
 \caption{Weak field regime ($|\mathcal B|<|\mathcal J|$): the entanglement measure $\mathcal{E}({\mathcal J},{\mathcal B},T)$ for a fixed value of $\mathcal J=-10$K and different values of $\mathcal B$.}
\label{FigBmenorJ} 
\end{figure}

At the weak field regime ($|\mathcal B|<|\mathcal J|$) -- with $\mathcal J=-10$K (see Fig. \ref{FigBmenorJ}) -- it can be verified that as temperature decreases the quantum entanglement measure increases monotonically and, as $|\mathcal B|$ approaches $|\mathcal J|$, at a given temperature, the quantum entanglement diminishes, besides that, as previously stated (\ref{TemCriDim2}), the decoherence temperature remains unaffected despite the presence of external magnetic field, therefore unaffected by the value of $|\mathcal B|$. It shall be stressed that since the system is an antiferromagnetic spin-1/2 dimer, the entanglement is insensitive in regard to the orientation of the external magnetic field along $z$-direction ($\hat{\rm\mathbf z}$ or $-\hat{\rm\mathbf z}$).  
 
\begin{figure}[h!]
\includegraphics[scale=0.5]{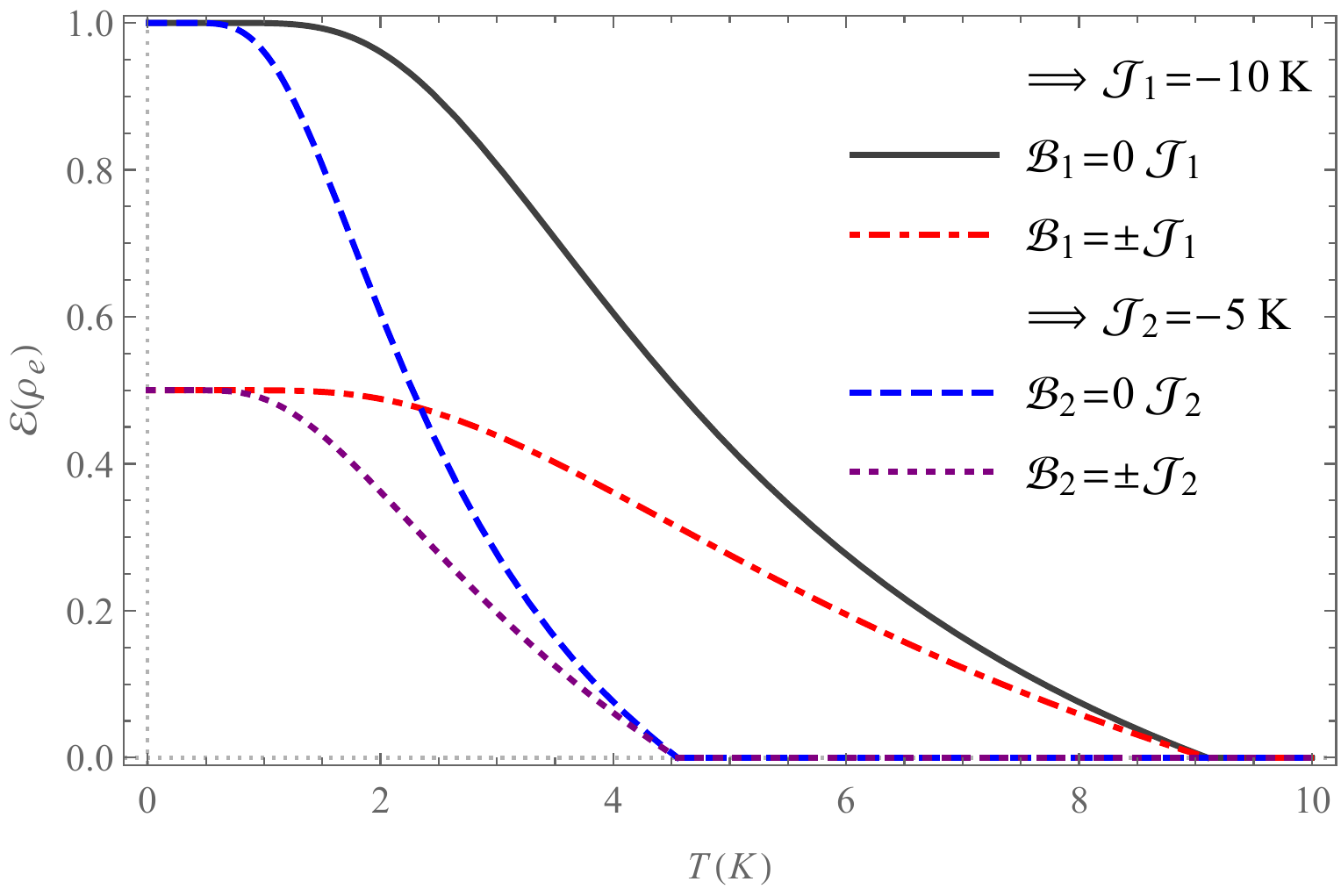}
 \caption{Medium field regime ($|\mathcal B|=|\mathcal J|$): the entanglement measure $\mathcal{E}({\mathcal J},{\mathcal B},T)$ for $\mathcal J=-5$K and $\mathcal J=-10$K, with $\mathcal B=0$ and $\mathcal B\neq0$.}
\label{FigBigualJ} 
\end{figure}

The entanglement behaviour at medium field regime ($|\mathcal B|=|\mathcal J|$) -- assuming $\mathcal J=-5$K and $\mathcal J=-10$K (see Fig. \ref{FigBigualJ}) -- is remarkable since no matter which the value of $\mathcal J$, as temperature approaches zero the maximum value of entanglement, $\mathcal E=1$, is reduced by half, $\mathcal E=1/2$, actually this result is valid whatever the values of $|\mathcal B|$ and 
$|\mathcal J|$, provided they are equal. It is also verified that the quantum entanglement measure increases monotonically as temperature decreases, as well as the critical temperature shows to be independent of applied magnetic field as given by Eq.(\ref{TemCriDim2}).
  
\begin{figure}[b!]
\includegraphics[scale=0.5]{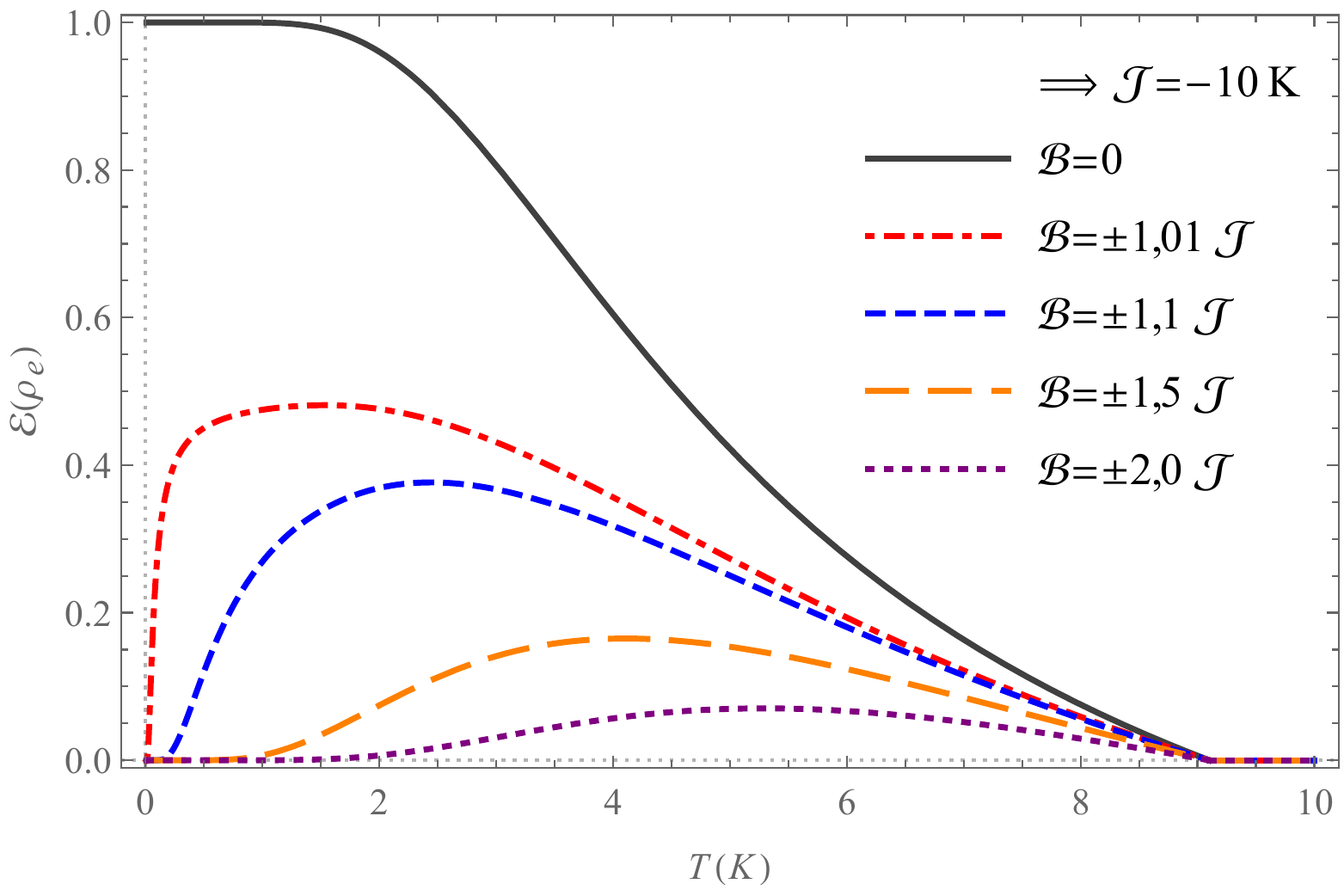}
 \caption{Strong field regime ($|\mathcal B|>|\mathcal J|$): the entanglement measure $\mathcal{E}({\mathcal J},{\mathcal B},T)$ for a fixed value of $\mathcal J=-10$K and different values of $\mathcal B$.}
\label{FigBmaiorJ}
\end{figure}

At the strong field regime ($|\mathcal B|>|\mathcal J|$) -- with $\mathcal J=-10$K (see Fig. \ref{FigBmaiorJ}) -- the quantum entanglement measure behaves differently if compared to the other both. In contrast to weak and medium regimes, where entanglement measure increases monotonically as temperature decreases, under strong magnetic field the quantum entanglement exhibits a maximum at a temperature, $T_{\rm m}$, such that, $0<T_{\rm m}<T_{\rm c}$, while the entanglement measure vanishes, no matter which the value of $|\mathcal B|$, when $T\rightarrow 0$ and $T=T_{\rm c}$.   

\subsection{Entanglement dependence on external magnetic field}

\begin{figure}[t!]
\includegraphics[scale=0.485]{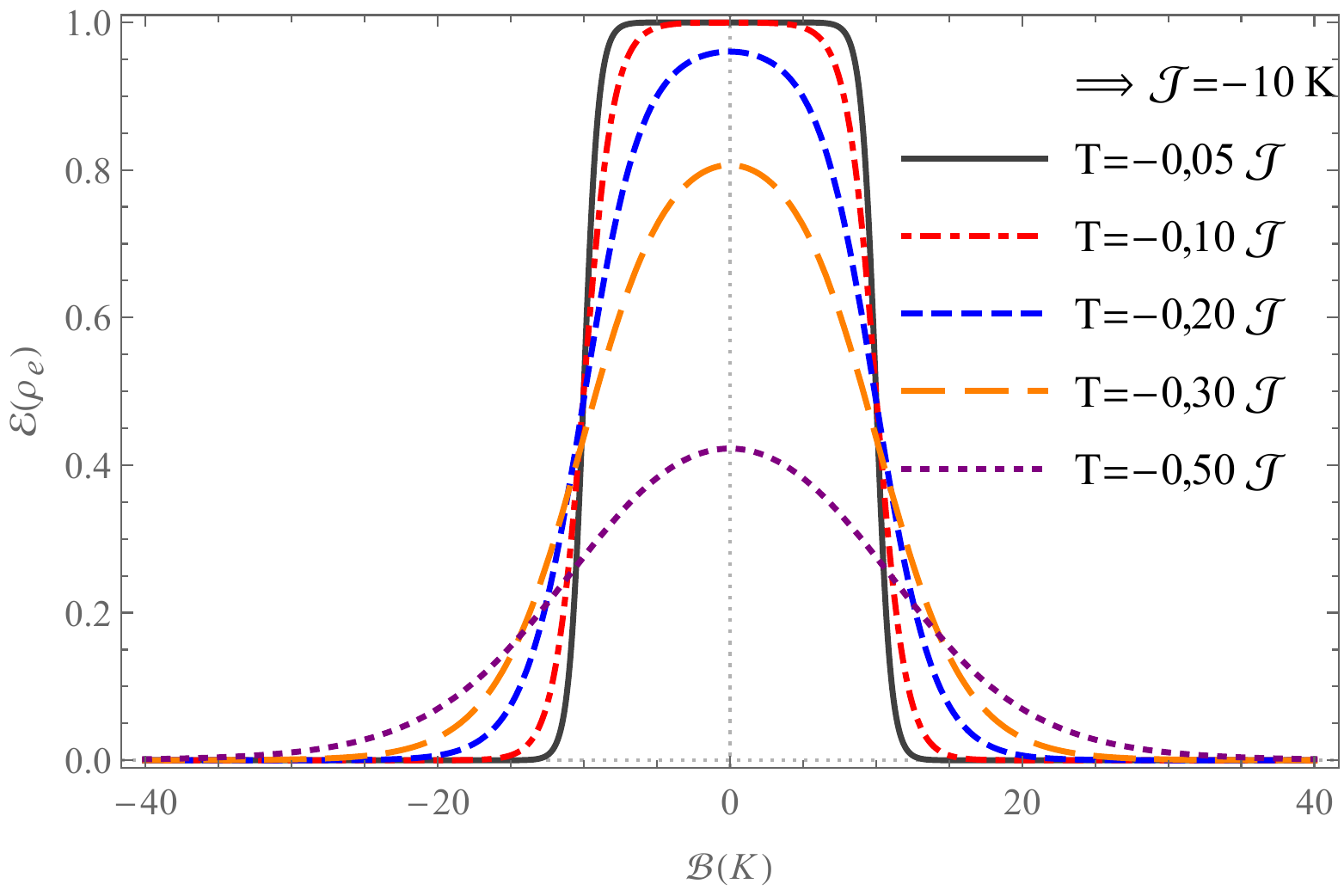}
 \caption{The entanglement measure $\mathcal{E}({\mathcal J},{\mathcal B},T)$ for a fixed value of $\mathcal J=-10$K and different values of $T$.}
\label{FigBeJFix} 
\end{figure}

Up to this moment it shall be concluded that the presence of external magnetic field affects the quantum entanglement, though the decoherence temperature lays undisturbed (\ref{TemCriDim2}). The behaviour of entanglement measure (\ref{RelEntDis}) of a particular material, namely for a given value of $\mathcal J$, at fixed temperatures but varying the external magnetic field, is presented -- with $\mathcal J=-10$K (see Fig. \ref{FigBeJFix}). As expected, and already mentioned, the entanglement depends only on the absolute value of the external magnetic field ($|\mathcal B|$) and not on its orientation along $z$-direction, because the system is a spin-1/2 antiferromagnetic dimer. However, it is foreseen that such a symmetry concerning the magnetic field ($z$-direction) orientation should be broken for mixed spin-(1/2,1) antiferromagnetic dimers \cite{spin-1/2-1}. It can be noticed that, from the entanglement measure curves as function of the applied magnetic field (see Fig. \ref{FigBeJFix}), there are an inflection point, whenever $\mathcal B =\mp\mathcal J$, which makes explicit the transition from weak to strong field regimes. Moreover, the abscissa axis is the horizontal asymptote of the entanglement measure as function of external magnetic field -- there is no $\mathcal B$ such that $\mathcal{E}$ vanishes -- namely, when $\mathcal B \rightarrow \pm\infty$ then $\mathcal E \rightarrow 0$. 

In order to illustrate the {\it quantum entanglement magnetic shielding effect} introduced in this work, a real system like [Fe$_2$(SC$_3$H$_5$N$_2$)$_2$(NO)$_4$], the binuclear nitrosyl iron complex \cite{nitrosyl}, is taken into account. The nitrosyl iron complex is a spin-1/2 antiferromagnetic dimer with $\mathcal J=-136$K, if it was subjected to an external magnetic field $|\mathcal B|= 10$K ($|B|\approx 7,456$T) at temperature $T=60$K, which is below the critical one, the entanglement measure would be $\mathcal E\approx 0,52$, as well as, if $|\mathcal B|= 140$K ($|B|\approx 104,238$T) then $\mathcal E\approx 0,32$. Therefore, the quantum entanglement phenomenon in spin-1/2 antiferromagnetic dimers survives even under the action of very high magnetic fields, like Tesla orders of magnitude. 

\begin{figure}[b!]
\includegraphics[scale=0.5]{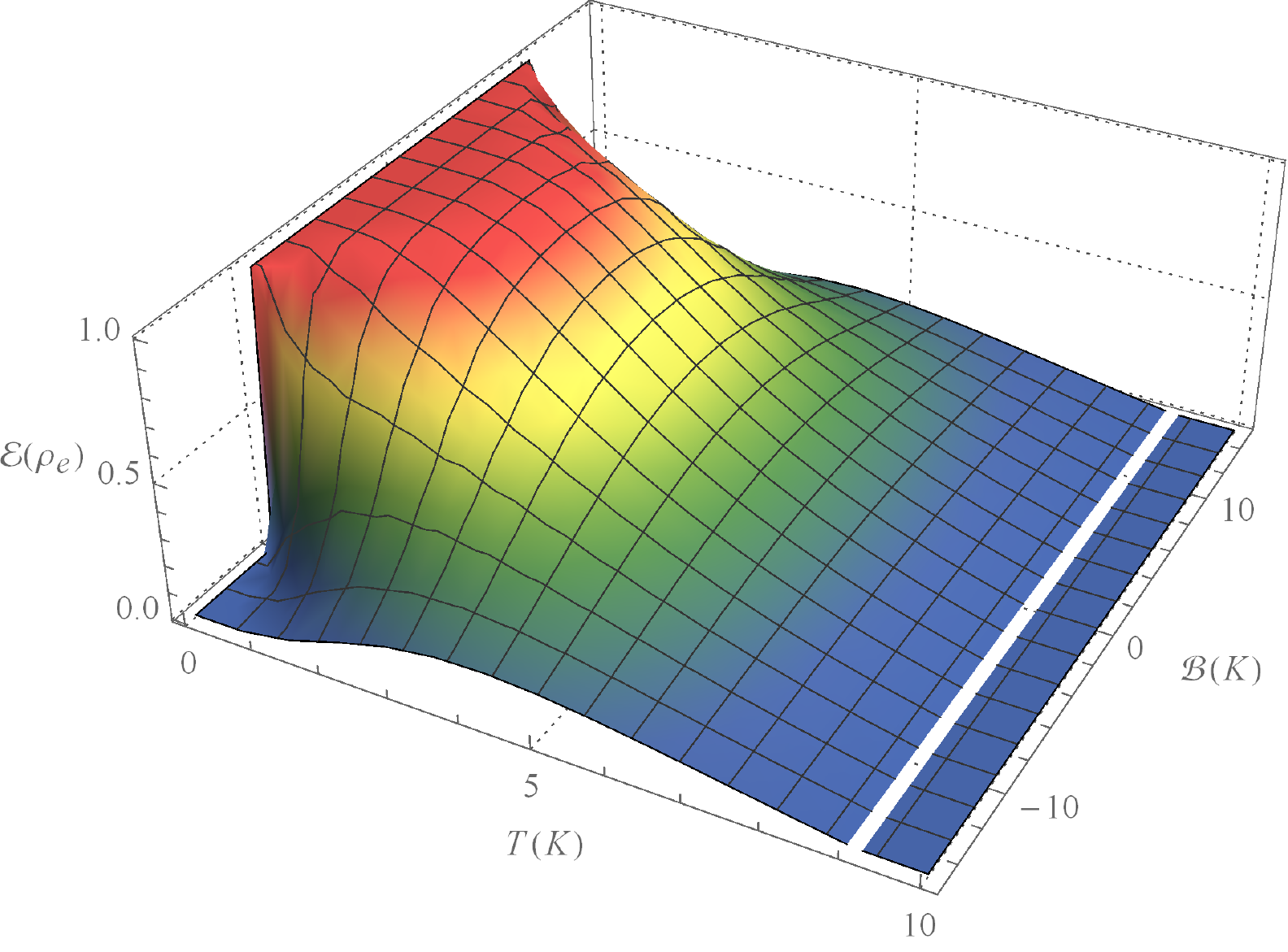}
 \caption{The entanglement measure $\mathcal{E}({\mathcal J},{\mathcal B},T)$ for a fixed value of $\mathcal J=-10$K.}
\label{Plot3D} 
\end{figure}

The entanglement measure (\ref{RelEntDis}) dependence on temperature and external magnetic field exhibits a straight line for a fixed value of temperature (see Fig. \ref{Plot3D}), thus parallel to the external magnetic field axis, where the entanglement vanishes, because $\rho_{23}^e=\rho_{23}^s$ owing to the fact that $T=T_{\rm c}$.

\section{Conclusions}
\label{conclusions}

To end, it follows the final conclusions and comments. The effect of temperature and external (applied) magnetic fields to quantum entanglement phenomenon on spin-1/2 Heisenberg dimer systems was investigated and analysed in details. It was proved that only antiferromagnetic exchange couplings shall favour quantum entangled states, thus there is no place for quantum entanglement in spin-1/2 ferromagnetic dimers. A physical quantity $\mathcal{E}({\mathcal J},{\mathcal B},T)$, expressed in terms of experimentally measurable quantities, was proposed so as to quantify the degree of quantum entanglement. Bore in mind the Peres-Horodecki criterion, the critical (decoherence) temperature, above which the entanglement disappears, 
$T_{\rm c}=-{\mathcal J}/\ln(3)$, was determined. The same result has been obtained by adopting the entanglement measure $\mathcal{E}({\mathcal J},{\mathcal B},T)$ which vanishes at $T=T_{\rm c}$, thus showing  that the entanglement measure proposed in this work consists in a good measure to quantify how entangled is a system at a given temperature and external magnetic field. In the following, it was discussed minutely the behaviour of the system from the quantum entanglement point of view with respect to temperature: weak field ($|\mathcal B|<|\mathcal J|$), medium field ($|\mathcal B|=|\mathcal J|$) and strong field ($|\mathcal B|>|\mathcal J|$). Further investigation should deserve attention to the contribution of anisotropy effects (impurities) to the degree of entanglement, critical temperature and its possible dependence on the external magnetic field, as well as an eventual origin of a critical decoherence magnetic field ($B_{\rm c}$). Analogously to what has been previously verified in spin-1/2 antiferromagnetic trimers and chains \cite{trimer-spin-1/2,chains-spin-1/2} -- the greater the $|{\mathcal J}|$, the greater is $T_{\rm c}$ -- it should also be conjectured for spin-1/2 antiferromagnetic dimers that the exchange coupling protects the system from decoherence as temperature increases. Moreover, as a remarkable by-product, it was shown that the critical temperature does not depend on applied magnetic field, as also pointed out in \cite{Gong2009,Mehran2014}, consequently, entangled states are somehow shielded even from the action of high external (Tesla orders of magnitude) magnetic fields -- the {\it quantum entanglement magnetic shielding effect} -- as presented for the case of the nitrosyl iron complex [Fe$_2$(SC$_3$H$_5$N$_2$)$_2$(NO)$_4$] \cite{nitrosyl}.

\subsection*{Acknowledgments} 

CAPES and FAPEMIG are acknowledged for invaluable financial help. The authors are grateful to A.R. Pereira, J.B.S. Mendes and L.G. Rizzi, as well as to the anonymous referee for helpful comments and suggestions. O.M.D.C. dedicates this work to his father (Oswaldo, {\it in memoriam}), mother (Victoria, {\it in memoriam}), daughter (Vittoria) and son (Enzo). M.M.S. dedicates this work to his grandmother (Geralda, {\it in memoriam}).



\end{document}